# Thermal properties of Rhea's Poles: Evidence for a Meter-Deep Unconsolidated Subsurface Layer

C.J.A. Howett[1], J.R. Spencer[1], T. Hurford[2], A. Verbiscer[3], M. Segura[2].

1 - Southwest Research Institute, Colorado, USA.

2 - Goddard Space Flight Center, Maryland, USA.

3 – University of Virginia, Charlottesville, Virginia, USA.

**Corresponding Author and their Contact Details:**

C.J.A. Howett

Email: howett@boulder.swri.edu

Telephone Number: +1 720 240 0120

Fax Number: +1 303-546-9687

Address:

1050 Walnut Street, Suite 300

Boulder, Colorado

80302

USA

**Abstract**

Cassini's Composite Infrared Spectrometer (CIRS) observed both of Rhea's polar regions during a close (2,000 km) flyby on 9th March 2013 during orbit 183. Rhea's southern pole was again observed during a more distant (51,000 km) flyby on 10th February 2015 during orbit 212. The results show Rhea's southern winter pole is one of the coldest places directly observed in our solar system: surface temperatures of 25.4±7.4 K and 24.7±6.8 K are inferred from orbit 183 and 212 data respectively. The surface temperature of the northern summer pole inferred from orbit 183 data is warmer: 66.6±0.6 K. Assuming the surface thermophysical properties of the two polar regions are comparable then these temperatures can be considered a summer and winter seasonal temperature constraint for the polar region. Orbit 183 will provide solar longitude ($L_s$) coverage at 133° and 313° for the summer and winter poles respectively, whilst orbit 212 provides an additional winter temperature constraint at $L_s$ 337°. Seasonal models with bolometric albedo values between 0.70 and 0.74 and thermal inertia values between 1 and 46 J m$^{-2}$ K$^{-1}$ s$^{-1/2}$ (otherwise known as MKS units) can provide adequate fits to these temperature constraints (assuming the winter temperature is an upper limit). Both these albedo and thermal inertia values agree within the uncertainties with those previously observed on both Rhea's leading and trailing hemispheres. Investigating the seasonal temperature change of Rhea's surface is particularly important, as the seasonal wave is sensitive to deeper surface temperatures (~tens of centimeters to meter depths) than the more commonly reported diurnal wave (typically less than a centimeter), the exact depth difference dependent upon the assumed surface properties. For example, if a surface porosity of 0.5 and thermal inertia of 25 MKS is assumed then the depth of the seasonal thermal wave is 76 cm, which is much deeper than the ~0.5 cm probed by diurnal studies of Rhea (Howett et al., 2010). The low thermal inertia derived here implies that Rhea's polar surfaces are highly porous even at great depths. Analysis of a CIRS focal plane 1 (10 to 600 cm$^{-1}$) stare observation, taken during the orbit 183 encounter between 16:22:33 and 16:23:26 UT centered on 71.7° W, 58.7° S provides the first analysis of a thermal emissivity spectrum on Rhea. The results show a flat emissivity spectrum with negligible emissivity features. A few possible explanations exist for this flat emissivity spectrum, but the most likely for Rhea is that the surface is both highly porous and composed of small particles (<~50 μm).

# 1 Introduction

On 9th March 2013 the Cassini spacecraft had a close (2,000 km) encounter with Saturn's mid-sized icy satellite Rhea. The remote sensing instruments onboard Cassini viewed Rhea's southern winter hemisphere on approach and Rhea's northern summer hemisphere during departure at approximately nadir geometry. Thus, during one encounter high-spatial resolution observations were obtained of both of Rhea's poles. Nearly two years later CIRS caught another glimpse of Rhea's southern pole, this time from further away (51,000 km) and at a high emission angle (80° to 90°). To date these data sets provide the best coverage of Rhea's polar regions by CIRS, as they were taken at high-spatial resolution and at mostly low emission angles.

The Composite Infrared Spectrometer (CIRS) is one of Cassini's remote sensing instruments and was taking data during both of these Rhea encounters. It is from these data that the surface temperatures of Rhea's polar regions can be inferred, where a polar region is loosely defined as lying between the pole and 60° N/S. CIRS has three focal planes covering 10 to 1400 cm$^{-1}$ (c.f. Flasar *et al.*, 2004). Focal plane 1 (FP1) covers 10 to 600 cm$^{-1}$ (16.7 to 1000 μm), enabling the temperatures of even very cold surfaces to be determined (<40 K). However, FP1's drawback is that it's made from a single circular detector, which has the lowest spatial resolution of CIRS' three focal planes (3.9 mrad/pixel). The other focal planes (focal planes 3 and 4, known as FP3 and FP4) cover 600 to 1100 cm$^{-1}$ (9.1 to 16.7 μm) and 1100 to 1400 cm$^{-1}$ (7.1 to 9.1 μm) respectively, and are both 1x10 arrays of 0.273 mrad/pixel detectors. These wavelength ranges make

FP3 and FP4 sensitive to surfaces warmer than 65 K and 110 K respectively, thus making them only suitable for looking at the warmer daytime temperatures of satellites and Enceladus' active south polar terrain.

In 2010 Cassini's Ion Neutral Mass Spectrometer (INMS) discovered traces of molecular oxygen and carbon dioxide around Rhea (Teolis *et al.*, 2010). Then evidence for a tenuous atmosphere around Dione was discovered using data collected in 2005 and 2010 by Cassini's Dual Technique Magnetometer (MAG) (Simon *et al.*, 2011). The composition of Dione's tenuous atmosphere was later shown (using 2010 Cassini Plasma Spectrometer data) to include molecular oxygen (Tokar *et al*, 2012). Follow up observations showed that the gas concentrations were higher over the northern hemispheres of both satellites (Teolis *et al.*, 2012), possibly due to seasonal variability. Since spring equinox in 2009 Rhea's (and Dione's) northern hemisphere has been warming, potentially causing volatiles previously trapped on its surface to sublimate. A surface capable of trapping the quantity of volatiles needed to produce the observed density of the exosphere must be both porous and cold (< 50 K) (Teolis *et al.*, 2010). Both of these requirements may be met in Rhea's polar regions. The presence of volatiles on the surface of Rhea, depending on their particle size and composition, could introduce observable emissivity variations into the CIRS spectrum. The previous study by Carvano *et al.* (2007) found no evidence of emissivity variations in the CIRS spectrum of Phoebe, Iapetus, Enceladus, Tethys or Hyperion, but did not consider Rhea or localized emissivity variations.

## 2 Data and Analysis

*2.1 Thermophysical Property Determination*

CIRS' FP1 captures the majority of the blackbody emission thus providing the most robust surface temperature determination for both colder and warmer surfaces. Table 1 shows the CIRS FP1 data obtained both during the March 2013 and February 2015 flyby that are analyzed in this work. The low altitude of the March 2013 Cassini flyby provides high spatial resolution observations even with FP1 (16 km/pixel to 359 km/pixel, see Table 1 for more information). Thus, during these two encounters FP1 has sufficient spatial resolution to determine polar temperatures and therefore only results from this focal plane will be discussed. During these observations the sub-solar latitude was low (18° N throughout the observation) so the northern sunlit pole was observed at high phase (72°). The more distant February 2015 observation had an FP1 spatial resolution between 194 to 202 km/pixel and viewed the southern pole at a high emission angle (>80°). Temperatures inferred from observations taken at high phase and high emission angles should be treated with a certain degree of caution because when rough terrain is viewed at these geometries non-representative surface regions are observed. For example during high phase observations a rough surface will create shadows that decrease the surface temperature observed by CIRS, which in turn would require a higher albedo and a lower thermal inertia to fit the observation. Whereas observations at high emission angles would preferentially observe high elevation regions (e.g. in this geometry equator-facing crater walls), the temperature of which may not be representative of the bulk surface temperature. These high elevation polar regions are probably warmer than their bulk surroundings since they will experience longer summer heating than the lower-lying and hence more shadowed neighboring terrain. Similar effects have already been observed on other icy satellites, for example Spencer (1987) analyzed Voyager 1 and 2's Infrared Interferometer Spectrometer and Radiometer (IRIS) observations of Callisto and found large variations of spectrum slope with emission angle at low solar elevation, which were attributed to the effect of surface topography. Such warmer temperatures require a lower albedo and higher thermal inertia to fit the observations.

For each CIRS spectrum, the best-fitting blackbody temperature emission curve is found using a downhill simplex method based on the work of Nelder and Mead, 1965, as implemented in the IDL "amoeba" routine. Where observations overlap we calculate the mean spectral radiance, find the blackbody spectrum that fits this mean spectrum, and assume the corresponding blackbody temperature for the overlapping region. The spectral noise is converted to temperature noise using a two-step Monte Carlo technique (as detailed below). These surface temperatures of Rhea are then mapped, and the results are shown in Figures 1, 2 (South and North polar regions as observed in orbit 183) and Figure 3 (South polar region as observed in orbit 212).

If we assume that surface thermophysical properties of both of Rhea's poles are the same, and that the temperatures of the fields of view that lie closest to the poles are comparable with those at the actual poles, then it is possible to use these observations to constrain Rhea's polar seasonal temperature variations. This is because both winter and summer temperature constraints are provided by these observations. There are of course potential problems with these assumptions. For example what if the thermophysical surface properties of each polar region greatly varies? That would mean that the surface temperatures close to the pole are in fact very different to those at the actual pole, and that the inferred temperatures are very dependent upon the size of the field of view (a problem since the FP1 field of view has a very different spatial resolution between the different polar observations). Another possible problem is if the two poles have very different thermophysical properties, which would undermine the conclusions drawn from our modeling.

It is difficult to test the validity of these assumptions because there is little in the published literature comparing the surface of Rhea's poles, primarily because of the lack of Cassini coverage. However, the recently produced PIA18438 global map of Rhea shows no significant color difference between the poles at extended optical wavelengths (0.930 μm, 0.568 μm and 0.338 μm). Additionally, Howett *et al.* (2014) showed that Rhea's thermal inertias (outside of the anomalous Inktomi crater region) are uniform across low- and mid-latitudes. The PIA18438 map reveals no Inktomi-like crater at high-latitudes on Rhea. Therefore there is nothing in the available color or surface property maps to imply that the thermophysical properties of Rhea's polar regions are likely to vary significantly, either across each polar region or between the poles. So in the absence of better temporal and spatial coverage it is reasonable to use these observations to provide the first opportunity to probe Rhea's seasonal temperature variation.

Seasonal temperatures are a uniquely useful quantity as it has a deeper skin-depth than diurnal temperature variations (i.e. they are sensitive to the deeper properties of the regolith). Thus, a combination of diurnal and seasonally derived surface thermophysical properties provides a strong constraint on how a surface regolith varies with depth. The depth the seasonal wave penetrates into the surface depends upon the thermophysical properties of the surface. Skin depth is defined as $\frac{I}{\rho c \sqrt{\omega}}$, where $\rho$ is the density, $c$ is the specific heat, $\omega$ is the angular velocity of rotation and $I$ is the thermal inertia (described as $\sqrt{k\rho c}$, or as $\sqrt{k\rho_0 (1-p)c}$), where $\rho_0$ is the zero porosity density, $p$ is the porosity and $k$ is the thermal conductivity). Thermal inertia and albedo are the two quantities this work aims to constrain, since they are the prime parameters controlling surface temperature. In essence thermal inertia describes how well a surface is able to store and release thermal energy, whilst the bolometric albedo $A$ describes the wavelength-integrated fraction of incident solar radiation reflected in all directions by a body's surface. (1-$A$) is thus the fraction of incident radiation that is absorbed and is available to heat the surface. To constrain these two thermophysical properties we compare the observed temperatures to the results of a 1-D diurnal thermophysical model (c.f. Spencer *et al.*, 1989) that has been modified to predict seasonal polar temperatures. The model solves for the one-dimensional heat flow conducted to and from the surface in order to calculate the temperature as a function of depth. The upper boundary is set so that thermal and incident solar radiation are balanced with the heat conducted to and from the surface, whereas the lower boundary is set deep enough that diurnal and seasonal temperature variations result in negligible temperature change at this level. The surface emissivity is set to unity, no additional geothermal heat flow is assumed. Solar insolation is calculated using Rhea's sub-solar latitude and heliocentric distance variations. The seasonal surface temperature variations, as a function of bolometric albedo and thermal inertia, are pre-calculated and stored as look-up tables.

Figures 4 and 5 show how the night and daytime temperatures observed by CIRS vary with latitude. The figures show that Rhea's South pole has a temperature of 25.4±7.4 K, and that the observation taken closest to Rhea's North pole has a temperature of 66.6 ±0.6 K. Note, a two-step Monte Carlo technique is used to convert the spectral noise estimate into a temperature noise. The process is fully described in Howett et al. (2011) but an overview is provided here for completeness. Firstly synthetic noise with a comparable magnitude to the observed noise is created and added to the previously determined best fitting blackbody curve. In the second step each spectrum (created in the first step) is fitted by a blackbody emission spectrum. These two steps are repeated numerous times and the temperature error estimate is given by the standard deviation of the temperatures whose blackbody emission spectra are best able to fit the created spectra.

Figure 5 also shows that the South pole temperature as inferred from the orbit 212 data is 24.7±6.8 K, which is slightly colder but agrees (within error) with the south polar temperature derived from orbit 183 data. These blackbody curves polar temperature are above, but close to, the limit CIRS can detect (compare the noise and signal in Figures 5b and 5c). So whilst the quoted errors are robustly derived these temperatures can conservatively be considered an upper-limit, which is how they're considered when used to constrain the seasonal model temperatures. Saturn's season is described by the solar longitude $L_S$, which is defined here to be 90° at the northern vernal equinox and 270° at the northern autumnal equinox. Thus, the orbit 183 north polar observations were taken at $L_S$ = 113° while the orbit 183 and 212 southern polar observations were taken at the equivalent $L_S$ = 313° and $L_S$ = 337° respectively (i.e. northern summer/ southern winter). If it is assumed that the observed southern pole (winter) temperatures are a proxy for those the north pole would experience at the same solar longitude then any credible seasonal model will have to predict winter and summer temperatures within the errors, at these $L_S$.

The seasonal model was run for albedo values between 0.60 and 0.80 in 0.01 increments and thermal inertias between 1 and 60 J m$^{-2}$ K$^{-1}$ s$^{-1/2}$ (henceforth referred to as MKS) in 1 MKS increments to produce the predicted seasonal temperature variations shown in Figure 6. As the figure shows, only albedo values of 0.72±0.02 and thermal inertias values of $24^{+22}_{-23}$ MKS are able to predict accurately the summer and winter seasonal polar temperatures observed on Rhea.

Additional seasonal model runs using an increasing thermal inertia profile with depth were also tested over a limited thermal inertia and albedo range (shown in Figure 7a). The depth profile used a nominal surface value for the top 1 cm, increasing to three times this nominal value at 3 cm, and five times its value by 10 cm (Figure 7b). Nominal surface thermal inertias between 1 and 50 MKS in 5 MKS increments, and albedo values between 0.69 and 0.75 in 0.02 increments were used. The only models able to fit the data had an albedo of 0.73 and a nominal surface thermal inertia of 1 or 5 MKS, increasing to a maximum thermal inertia of 10 MKS and 50 MKS respectively (Figure 7c). The albedo and thermal inertia of these model fits is in good agreement with those of the single thermal inertia model fits: the albedos agree within error, and only the highest thermal inertia observed at large skin depths (50 MKS below 10 cm) are just outside the range previously derived.

*2.2 Emissivity Determination*

Since emissivity variations are subtle, high signal-to-noise observations are required to constrain them. Co-adding observations taken whilst the CIRS field of view remains at a single location (known as stare observations) increases the signal-to-noise. Volatile cold-traps are most likely to occur on the coldest surfaces of Rhea: this is the winter South Pole at the time the data was acquired. CIRS obtained a separate set of 12 stare observations near this region (centered at 71.7º W, 58.7º S) during orbit 183 on March 9$^{th}$ 2013 between 16:22:33 and 16:23:26 UTC. Table 1 shows full details of the geometry of the observation during this time. The location of the FP1 field of view is given in Figure 8(a). The spectra obtained during this stare are shown in Figure 8(b), along with the mean value and best fitting multi-component blackbody spectrum. In order to obtain the best-fitting spectrum we allow a combination of two blackbody spectra assuming a unity bolometric emissivity. The best fit to the observed spectrum is obtained assuming a surface temperature of 54.9±13.6 K over 24.8±17.7 % of the field of view, with the remainder at 31.4±10.5 K. This result implies large temperature contrasts in the south-polar region, with much of the surface being extremely cold. The stare observations were taken at high solar incidence angles: 61º (sun is low on the horizon) to 100º (sun is below the horizon). If the surface is not smooth then under these illumination conditions (up to incidence angles of 90º) large shadows would be expected, which may somewhat explain the observed temperature contrasts. Whether shadowing effects alone can explain such large temperature contrasts warrants further attention, being beyond the scope of this work.

The standard deviation of the stare observations is also given in Figure 8(b), providing an estimate of the random noise in the data. Figure 8(c) shows the deep space spectrum taken close in time to the observation and can be used to provide an estimate of potential CIRS FP1 systematic errors. As the figure shows each of the five deep space spectrum have a very low (~negligible) magnitude that averages to zero within the errors and shows no significant errors except for wavenumbers smaller than 60 cm$^{-1}$. The ratio between the observed spectrum and the best fit gives an estimate of the spectral emissivity shown in Figure 8(d) with error bars showing the standard error of the mean for both the random noise and systematic errors. Finally, Figure 8(e) shows the brightness temperature of the mean radiance spectra given in Figure 8(b). If all of the observed surface were radiating at a single temperature then the brightness temperature would be constant with respect to wavelength. Instead the brightness temperature increases with wavenumber, which instead implies a range of temperatures were observed (e.g. Bandfield *et al.*, 2015). It's possible that these temperature variations are due to topographic/roughness variations within the CIRS field of view, similar to those observed by IRIS on Callisto (Spencer, 1987).

## 3 Discussion

The temperature observed at Rhea's winter pole matches the coldest surface temperature ever directly measured in our solar system, within the Hermite Crater at the lunar pole (Paige *et al.*, 2010). Figure 9 compares this temperature to those of other cold Solar System bodies. Bolometric albedo values between 0.70 and 0.74 are able to provide seasonal model fits to the observed winter and summertime polar temperatures on Rhea (Figure 6). These albedos are at the upper end but in keeping with those previously determined for Rhea's trailing and leading hemispheres (Howett *et al.*, 2010, Figure 9), especially since they represent somewhat of an upper-limit (see the earlier discussion on shadowing effects).

The thermal inertias able to fit the observations lie between 4 and 46 MKS, which encompasses the range of thermal inertias derived for Rhea's trailing and leading hemispheres using diurnal temperature observations: $8^{+12}_{-5}$ and $9^{+9}_{-5}$ (Howett *et al.*, 2010). The skin depth probed by the seasonal (and diurnal) waves, as previously described, is directly proportional to thermal inertia and the square root of the angular frequency (i.e. the diurnal or seasonal cycle). Assuming a surface density of non-porous ice at 93 K of 0.934 g cm$^{-3}$, a specific heat for water ice at 90K of 0.8 K J$^{-1}$ g$^{-1}$ (Spencer and Moore, 1992), a 0.5 porosity, a diurnal cycle of 4.518 Earth days and a thermal inertia of 8 MKS (Howett *et al.*, 2010) then Rhea's diurnal wave skin depth is ~0.6 cm. Assuming the same water ice density, specific heat and porosity but a seasonal cycle of 10832 Earth days and the range of seasonal thermal inertias derived here then the seasonal skin depth is between 12 and 140 cm. Thus, assuming the same porosity the seasonal wave is probing Rhea's surface 20 to >200 times deeper than the diurnal wave. However, Rhea's surface porosity is not well constrained. Laboratory studies estimate any icy satellite body with a diameter approximately less than or equal to Rhea's should have a residual porosity of >0.3 (Yasui and Arakawa, 2009). So instead the seasonal wave skindepth is better expressed as $1.53I/(1-p)$ cm, where p is the porosity and I is the thermal inertia in MKS. For comparison this is ~49 times larger than the diurnal skindepth: $0.031I/(1-p)$ cm, due to the large difference in the diurnal and seasonal timescales. To further illustrate this point Figure 11 shows how the seasonal skin depth varies with porosity for the mean polar thermal inertia and its limits. The figure shows that the skindepth of the seasonal wave is between 52 and 92 cm for a 24 MKS surface (for porosities between 0.3 and 0.6), and between ~2 and 175 cm over the full thermal inertia range (1 to 46 MKS). These depths probed by the seasonal wave are much deeper than previously probed by CIRS diurnal observations of Rhea and all of Saturn's classical icy satellites (see Figure 12).

The upper boundary of these thermal inertia values is surprisingly low. It might be expected that the ice surface would compact quickly with depth, increasing the thermal inertia. However, the upper limit of the polar thermal inertia value is much smaller than seen on the upper surface of the Galilean satellites (50 to 70 MKS, see references inside Howett *et al.*, 2010) and inside the thermal anomaly on Mimas (66 MKS, Howett *et al.*, 2011). It is also consistent with the seasonal thermal inertia upper-limit set for Enceladus

North polar region: < 100 MKS (Spencer *et al.*, 2006). In short, these results are consistent with Rhea's surface being fluffy to depths of tens of centimeters.

Figure 8(d) shows the emissivity variation observed for the FP1 stare observation, the location of which is shown in Figure 8(a). Figure 8(d) shows that the CIRS emissivity is unity at the 1σ level across the highest signal-to-noise part of the spectrum (~20 to 200 cm$^{-1}$; 50 to 500 μm). Outside of this region the CIRS emissivity is much noisier, but does not deviate from unity within 1σ, with the exception of the ~230 to 240 cm$^{-1}$ wavenumber region. However, it is not believed that this wavenumber region contains a real emissivity feature, as its >1 value is non-physical and its deviation from unity is still <2σ. Additional stare observations taken over a longer duration (therefore providing a higher signal-to-noise) are necessary to unambiguously resolve this issue, and detect any potentially weak emissivity signal that is currently lost in the noise. Thus, we conclude that no significant deviations from unity are observed in the emissivity spectrum, either as a slope or discrete feature, and it can be considered uniform.

Carvano *et al.*, (2007) modeled the emissivity of various grain sizes of water ice (normalized at 200 cm$^{-1}$) and showed that large grain sizes (500 and 1000 μm) have approximately unity emissivity over ~60 to 300 cm$^{-1}$, whilst that of smaller grain sizes (50 to 100 μm) have less than unity emissivity both at wavenumbers less than ~150 cm$^{-1}$ and greater than ~250 cm$^{-1}$. The same study also showed small grains (50 μm) of tholin material have ~10% decreasing emissivity between ~60 to 300 cm$^{-1}$, while larger grains (>500 μm) have a unity emissivity over the same wavenumber range. Spilker et al. (2005)

showed Saturn's A, B and C rings display a ~30% increase in emissivity between 20 and 60 cm$^{-1}$, leveling off at higher wavenumbers at emissivities between 0.75 and 0.85. The CIRS emissivity spectrum does not display any such trends in emissivity.

Emissivity variations occur when the observed surface does not radiate as a blackbody, which may occur for many reasons and can be manifest as slopes or discrete features. For example if a rough surface is remotely observed at high solar incidence angles then it will view both shadowed/non-shadowed regions, which will likely have vastly different surface temperatures. This will cause a range of surface emission, which when folded together may produce a spectrum that deviates from a single blackbody, producing a slope in the emissivity (Davidsson *et al.*, 2015; Bandfield *et al.*, 2015). Similarly spatial variations in the composition or thermal inertia of the surface can also lead to emissivity variations. Conversely surfaces made up of large particles (>500 *μm*) and surfaces made of small particles (i.e. sizes smaller than the observing wavelength, which in the high signal-to-noise region of FP1 is <50 μm) with large porosities (>90%) can produce a featureless emissivity spectra over the FP1 wavenumber range (Carvano *et al*., 2007). A final, but less likely option is that a spectrally featureless contaminant can suppress any volatile emissivity features.

Carvano et al. (2007) showed such feature suppression requires the surface to consist of >40% contaminant, which is incompatible with observations made by Cassini's Visual and Infrared Mapping Spectrometer (VIMS). Ciarniello *et al*. (2011) showed Rhea's water ice is very pure, with models only requiring <0.4% Triton tholin-like contaminants to fit the data. Additionally, this study found Rhea's polar regions to have a high albedo, between 0.70 and 0.74, which implies there is no (or at least very little) dark material contamination. Analysis of observations made across Rhea by Cassini's Visual and Infrared Mapping Spectrometer (VIMS) shows grain size across Rhea to be of the order of tens and not hundred of microns (e.g. Filacchione *et al*., 2012, Ciarniello *et al.,* 2011; Scipioni *et al.,* 2014). Hapke modeling has already shown that icy satellite surfaces have high porosity, for example porosity values between 65 and 95% have been calculated for Rhea (Domingue *et al*., 1995; Ciarniello *et al.,* 2011). Furthermore, the low thermal inertia limit determined in this work (≤46 MKS) is consistent with a porous surface, even at seasonal wave depths. Thus, a surface made up of small particles with large porosities is the most likely explanation for the flat emissivity spectrum observed. However, other explanations are possible, for example surface roughness effects maybe masking actual slopes in the emissivity (e.g. 50 μm tholins, as described by Carvano et al., 2007).

The original motivation to investigate this particular observation was to search for evidence of volatiles located in cold-traps in Rhea's polar regions. However, no spectral signatures were observed in the small region investigated close Rhea's north pole. If the bland emissivity spectrum were caused by a highly porous surface composed of small grains then this region would allow effective adsorption (Ayotte *et al*., 2001). However, since no volatile spectral signature was observed this implies that either the volatiles are located at depths CIRS is insensitive to (Figure 11 and 12), or that the volatile concentration (and hence signature) is too low in this region for CIRS to detect. Both options are feasible as CIRS would not be sensitive to any volatiles condensing on the top few mm of the surface and we know from observations of Rhea's atmosphere that the available volatile concentration is low. In fact Rhea's exosphere is very tenuous: it is two orders of magnitude lower than Europa's or Callisto's. A mean atmospheric $O_2$ column density of $3.4 \pm 0.7 \times 10^{16}$ m$^{-2}$ was inferred from Cassini/INMS measurements (the $CO_2$ density is lower still), which is well below the $10^{18}$ m$^{-2}$ lower detection limits of both Cassini's Magnetospheric Imaging Instrument (MIMI) and VIMS (Teolis *et al.*, 2010).

## 4 Conclusion

The derivation of thermophysical properties of Rhea's polar region over seasonal timescales shows that Rhea's polar surfaces do not compact quickly with depth, but rather they remain porous to depths of possibly several meters. One implication is that Rhea's polar surfaces (and perhaps by extension much of Rhea) are more efficient than expected at adsorbing volatiles. However, no evidence for surface volatiles was found in the emissivity spectrum of the south polar region. This could be explained by a combination of a polar surface that has a low porosity and made up of small particles. In this study CIRS data was used to determine the emissivity of only one specific region at the southern polar region. Thus, this result does not rule out the occurrence of further emissivity variations in other CIRS observations of Rhea.

**Tables**

| Date | Observation Details | Start Time (UT) | End Time (UT) | Sub-spacecraft longitude (° W) | Sub-spacecraft latitude (°) | Range (km) | Sub-Solar longitude (° W) | Spatial Resolution of FP1 (km/pixel) |
|---|---|---|---|---|---|---|---|---|
| 9$^{th}$ March 2013 | FP1 Map of Rhea's South Polar Region | 15:58:14 | 16:22:23 | 126° W to 128° W | 64° S | 63,203 to 77,329 | 11.0° W to 12.4° W | 246 to 301 |
| 9$^{th}$ March 2013 | FP1 stare centered at 71.7° W and 58.7 ° S | 16:22:33 | 16:23:26 | 128° W | 64° S | 62,621 to 63,110 | 12.4° W to 12.5° W | 244 to 246 |
| 9$^{th}$ March 2013 | FP1 Scan across Rhea's North Polar Region | 18:20:11 | 18:25:14 | 197° W to 251° W | 56° N to 73° N | 4,168 to 4,678 | 18.8° W to 19.1 ° W | 16 to 18 |
| 10$^{th}$ February 2015 | FP1 Scan from equatorial latitudes to Rhea's South Pole | 07:38:00 | 07:55:00 | 304° W to 310° W | 24° S to 27° S | 49,744 to 51,773 | 183 ° W to 184° W | 194 to 202 |

Table 1 – Details of the encounter geometry during various significant periods of the March 9$^{th}$ 2013 and February 10$^{th}$ 2015 Cassini Rhea flybys. The sub-solar latitude was 18° N and 23.6° N during the entire observing period of the two flybys respectively.

**Figures**

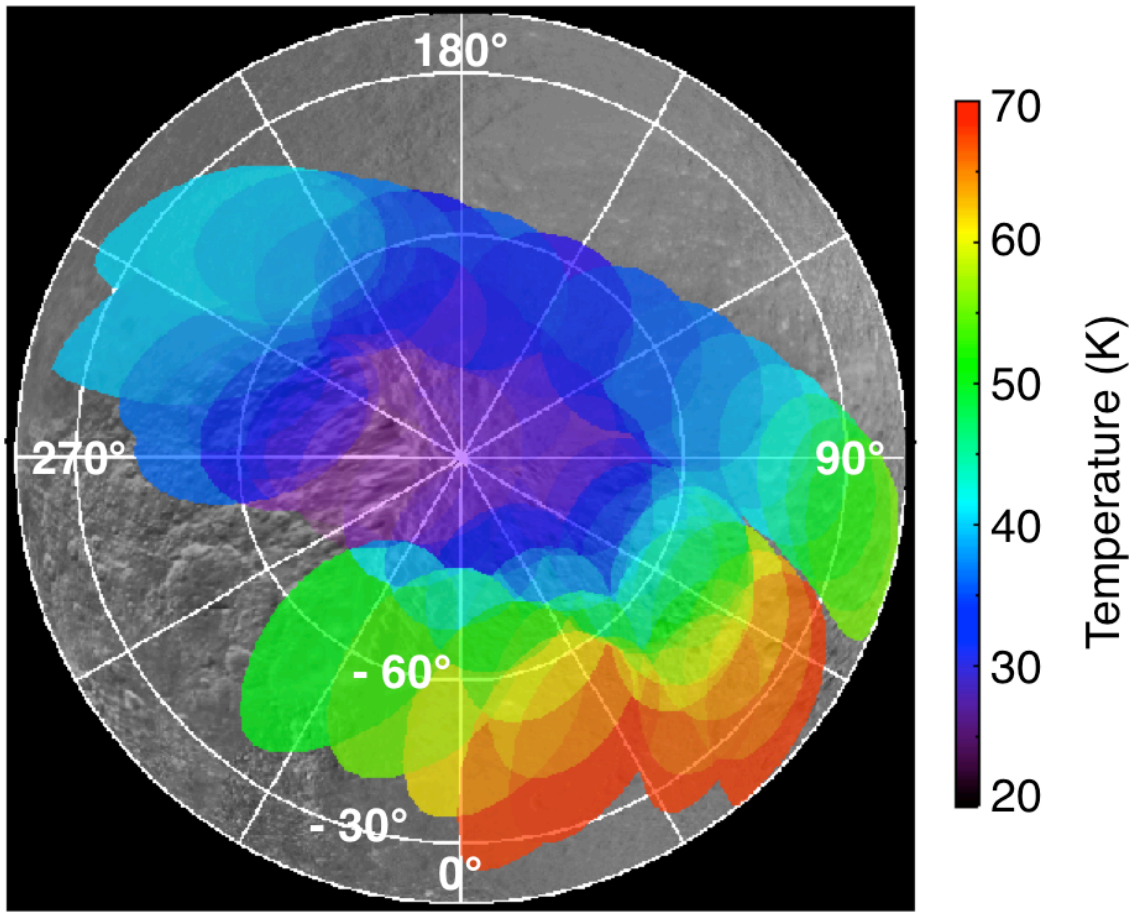

Figure 1 – Orthographic projection map centered on Rhea's South Pole, showing the orbit 183 observed surface temperature. The sub-solar point changes from 11° W, 18° N to 12° W, 18° N during this observation period. The base-map is Rhea ISS map PIA08343.

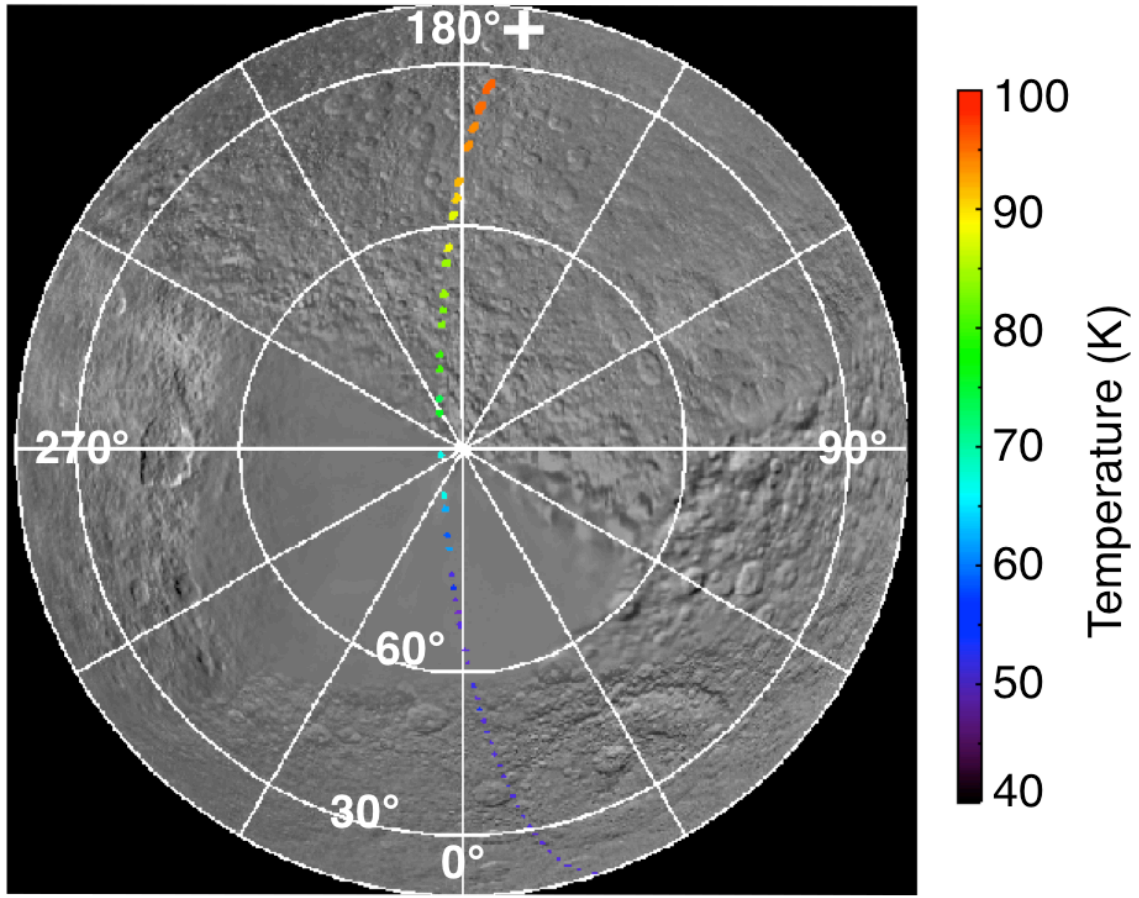

Figure 2 – Orthographic projection map centered on Rhea's North Pole, showing the surface temperatures derived from orbit 183 observations. The white cross indicates the sub-solar point; the base-map is Rhea ISS map PIA08343.

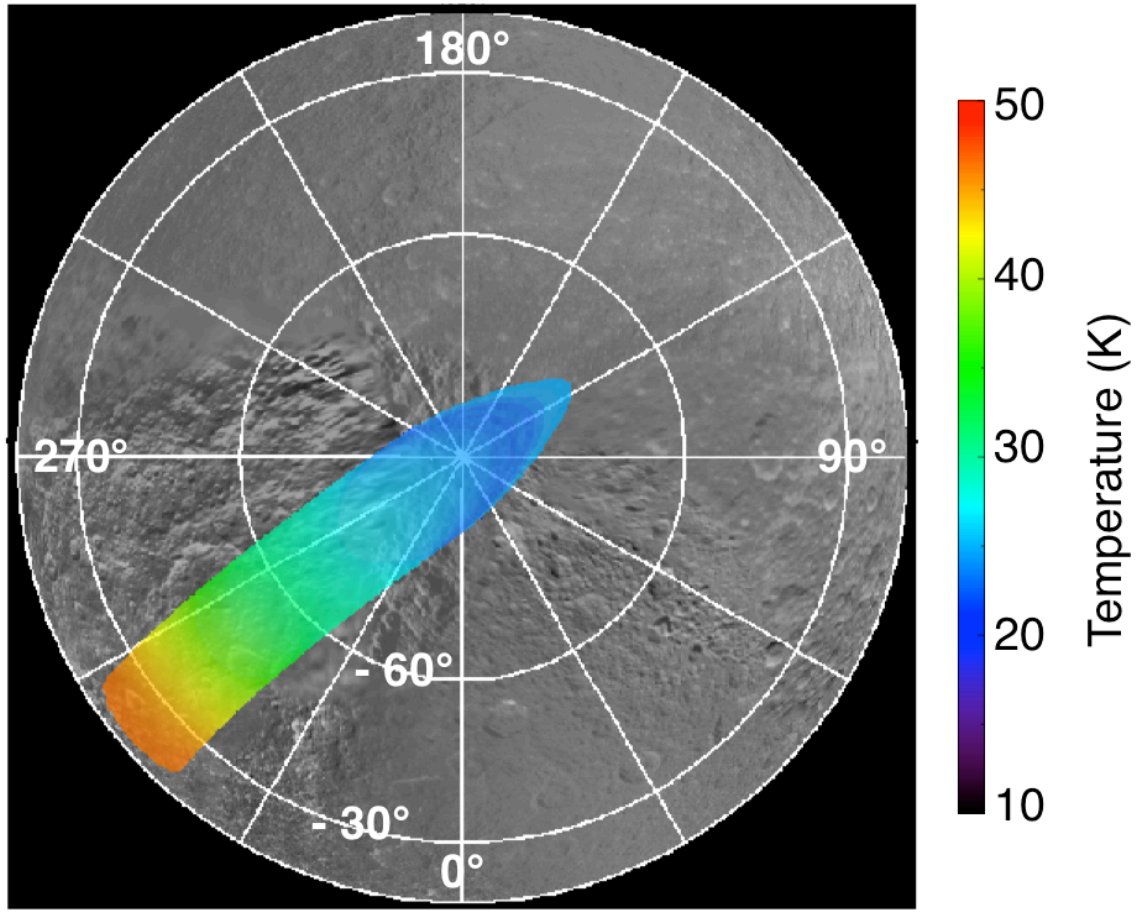

Figure 3 – Orthographic projection map centered on Rhea's South Pole, showing the orbit 212 observed surface temperature. The sub-solar point changes from 183° W, 24° N to 184° W, 24° N during this observation period. The base-map is Rhea ISS map PIA08343.

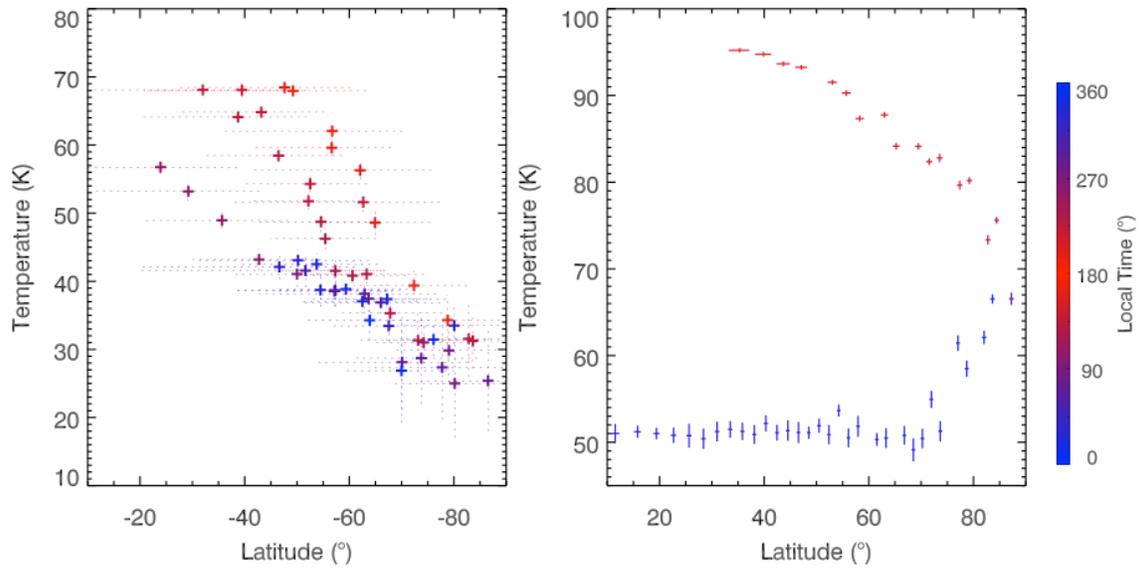

Figure 4 – Temperature variations with latitude for Rhea's orbit 183 South polar observation (left) and North polar observation (right). Daytime (sunlit) observations are in red; nighttime observations are in blue. Local time is defined as 90° at dawn and 270° at dusk.

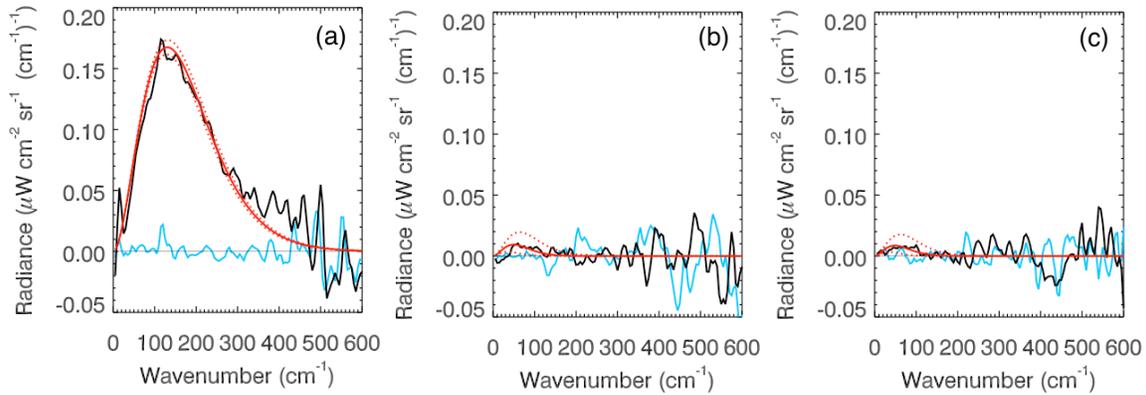

Figure 5 – CIRS FP1 radiance measurements of surfaces nearest to Rhea's North pole from orbit 183 (a) and South pole from orbit 183 (b) and orbit 212 (c). The red solid line indicates the best blackbody fit to the observed radiances, with the uncertainty indicated by the red-dotted line. The best fit temperature for the North pole observation is 66.6±0.6 K, whilst the South pole temperatures from orbit 183 and 212 are 25.4±7.4 K and 24.7±6.8 K respectively. The blue lines show the deep-space spectrum obtained closest to the observation time, providing a measure of the uncertainty. If the instrument and calibration were perfect the deep-space spectrum would be zero. The south pole detection matches the coldest directly detected temperature on a Solar System body.

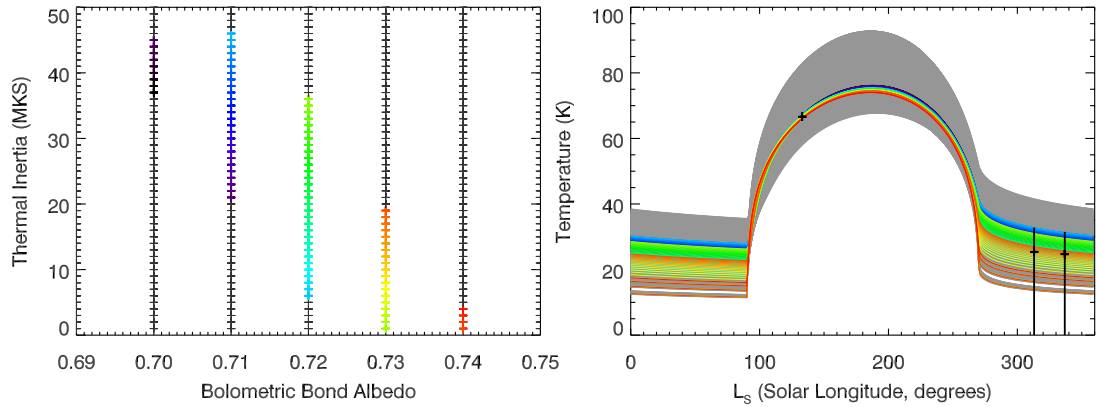

Figure 6 – The grey and colored crosses on the left show the range of albedo and thermal inertia combinations used by the model to predict the seasonal polar temperatures variations, which are shown on the right. The black crosses on the seasonal temperature curves show the observed summer and winter polar temperatures, as deduced in this work. The seasonal curves able to predict both the summer and winter temperatures appear in color, and the combination of albedo and thermal inertia values used to produce them are shown in the corresponding color in the left-hand figure. Assuming that the thermophysical properties of the two poles are similar the southern pole temperatures are used as the winter temperature constraint (Ls=313°), whilst the northern pole temperatures are shown at the actual encounter solar longitude (Ls=133°).

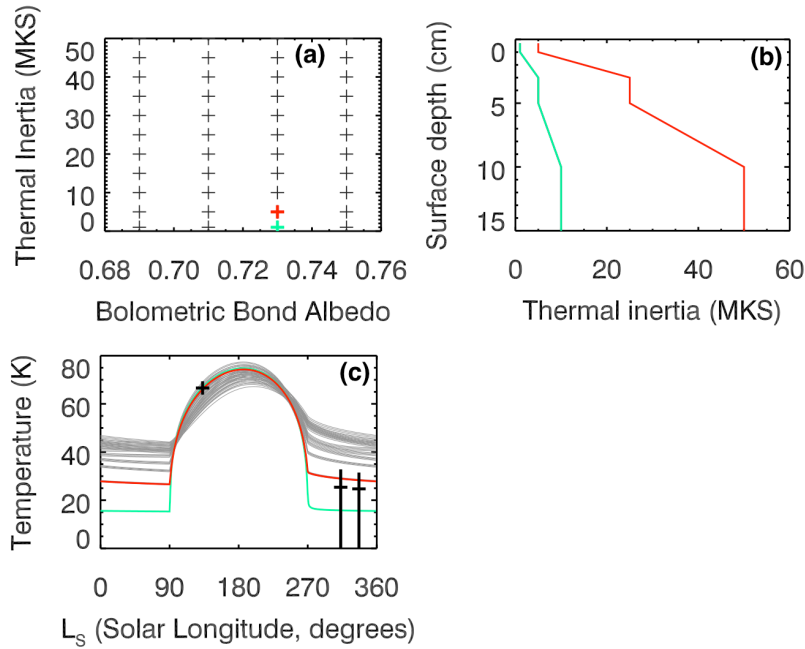

Figure 7 – Seasonal model fits to the observed polar temperature of Rhea using a thermal inertia profile that increases with depth. (a) The range of albedo and surface thermal inertia values tested by the model are shown by the crosses. Only those shown by the bold turquoise and red crosses were able to fit the observed temperatures. (b) The thermal inertia with depth profile for the two models able to fit the observed temperatures, shown in the corresponding color. Note the modeled thermal inertia remains constant below 3 cm extending to ~5 m in depth, but only the top 15 cm are show for clarity. (c) All of the model seasonal temperatures are shown (grey and colored curves). Rhea's observed polar temperatures and their errors are shown by the thick black crosses, using the same definitions as previously described in Figure 6. The two models able to fit the observed temperatures are shown in corresponding colors to their model parameters given in (a).

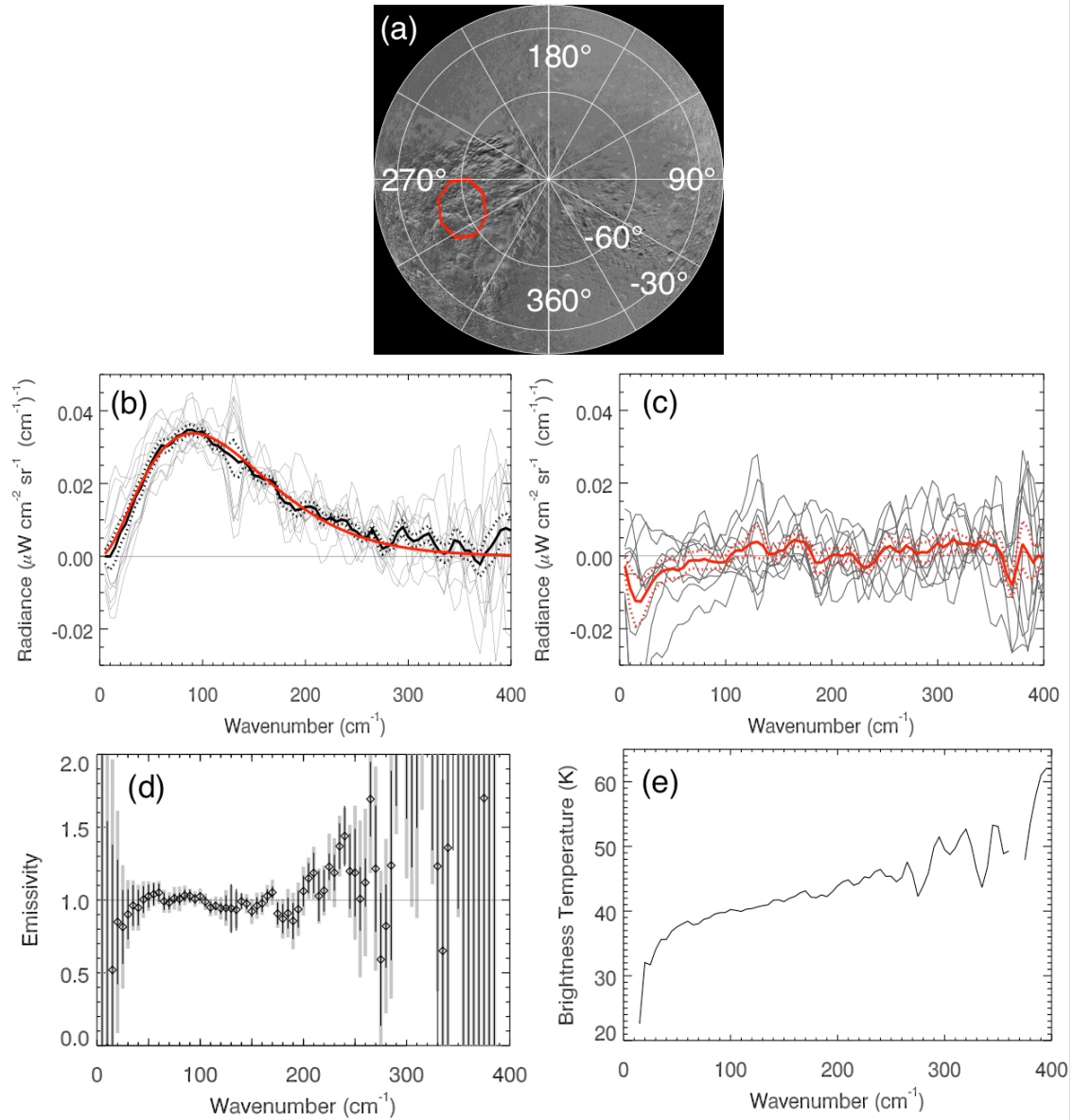

Figure 8 – (a) The location of the CIRS south polar stare at Rhea between 16:22:33 and 16:23:26 UTC. (b) The 12 spectra obtained during the stare observation (grey), the mean spectrum and the uncertainty in the stare observations are shown by the black solid and dotted lines respectively. Overplotted in red is the best fitting multi-component blackbody spectrum to the mean radiance (54.9±13.6 K over 24.8±17.7 % of the field of view, 31.4±10.5 K over the remainder) (c) The five deep space spectra taken before and after the stare observations are shown in grey, they are plotted on the same radiance scale as the stare observations shown in (b). The mean of these deep space spectra and its standard error are given by the red solid and dotted lines respectively. The deep space observations provide a measure of the systematic errors in the CIRS data, and show no significant errors except for wavenumbers smaller than 60 cm$^{-1}$. (d) The emissivity variation with wavenumber of the stare observation. The solid and dotted black error bars give the 1- and 3-sigma random noise error estimate determined from the variation in the stare observations respectively, whilst the light grey error bars provide an estimate of systematic error provided by the observed variation in the nearby deep space spectra. No believed emissivity variations above the noise levels are observed from unity, see main text for details. (e) The brightness temperature of the mean spectra, shown by the red colored line in Figure 7(b).

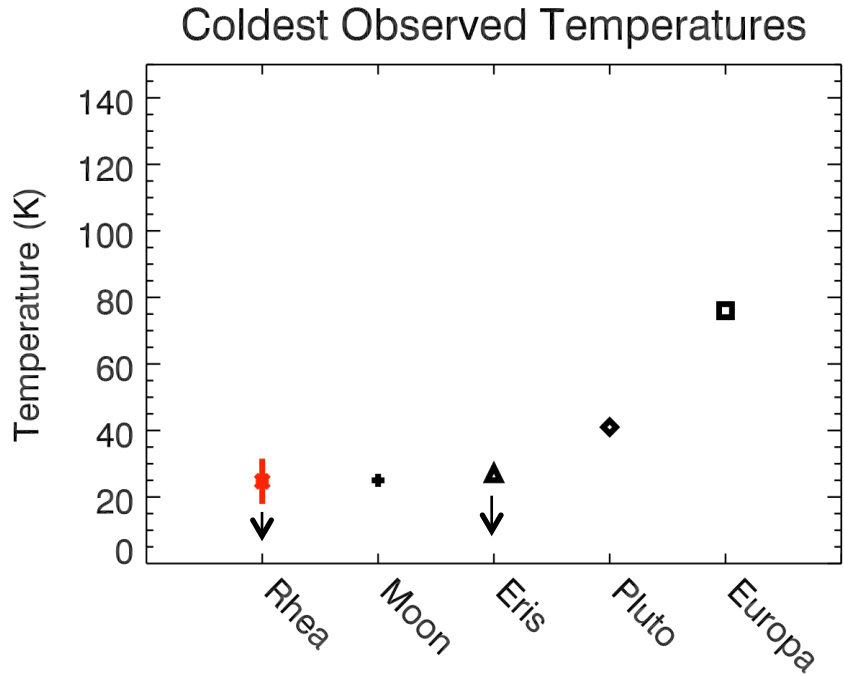

Figure 9 – Comparison of the coldest of Rhea's winter polar temperatures observed in this work: 24.7±6.8 K from orbit 212 data, which is considered an upper-limit in modeling work (indicated by the downward facing arrow), to that of directly observed surface temperatures of other cold bodies in our solar system. The coldest Lunar temperature was observed by Lunar Diviner on Lunar Reconnaissance Orbiter in Hermite Crater by Paige *et al*., (2010). This Eris surface temperature is an upper-limit, it was determined using 300 to 2480 nm ESO-Very Large Telescope observations by Alvarez-Candal *et al.* (2011). Pluto's surface temperature was determined from 1.4 mm Spitzer observations, from which the brightness temperature was converted to a physical (kinetic) temperature by assuming an emissivity of 0.9 (Lellouch *et al*., 2011). Please note the given surface temperature for Europa is actually a brightness temperature, it was observed at 70° S / 240° W by the photopolarimeter-radiometer onboard Galileo using 0.35 to ~100 μm data (Spencer *et al*., 1999) and thus the physical temperature at this location is likely to be higher. It is predicted that Europa's polar temperatures get as low at 50 K (Prockter and Pappalardo, 2006).

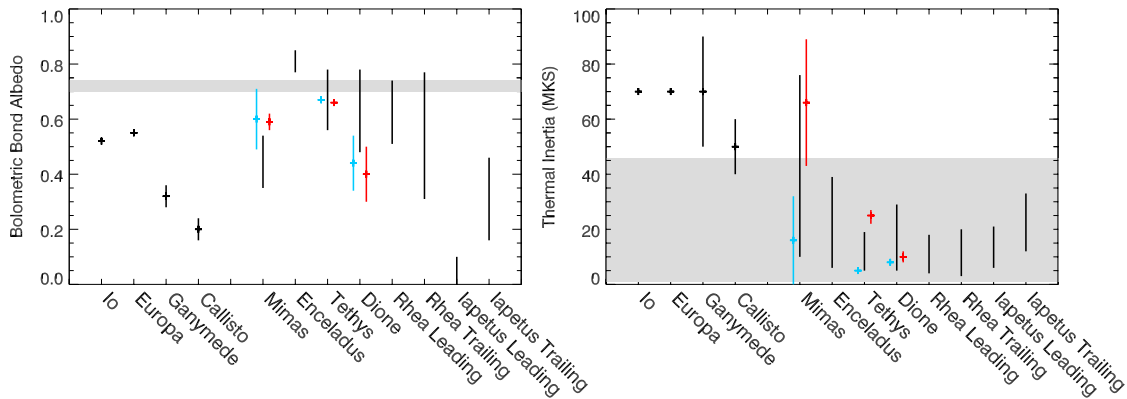

Figure 10 – The range of thermal inertia and bolometric albedo values able to fit the observed seasonal temperatures of Rhea's poles are shown by the grey shading. These thermal inertias and albedos are compared to those seen on other icy Saturnian satellites (black lines) (Howett *et al*., 2010), as well as those inside (red) and outside (blue) the thermally anomalous regions observed on some of these bodies (Howett *et al*., 2011, 2012, 2014). For context, these values are also compared to the Jovian Galilean satellites (Rathbun *et al*., 2010, Rathbun *et* al., 2004, Simonelli *et al*. (2001), Spencer *et al*., 1987).

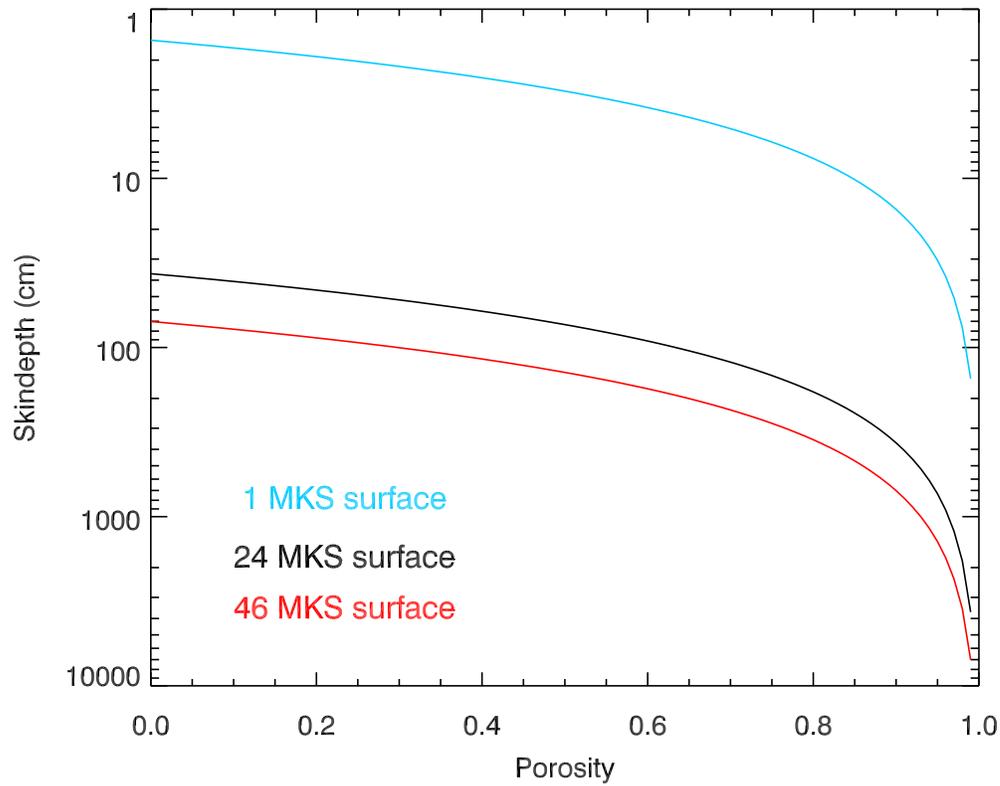

Figure 11 – The variation in seasonal wave skindepth with porosity for the mean (24 MKS) and lower/upper thermal inertia limits determined for Rhea's polar regions. A zero porosity surface density of 1.0 g cm$^{-3}$, a specific heat for water ice at 90 K of 0.8 J K$^{-1}$ g$^{-1}$ (Spencer and Moore, 1992) is assumed.

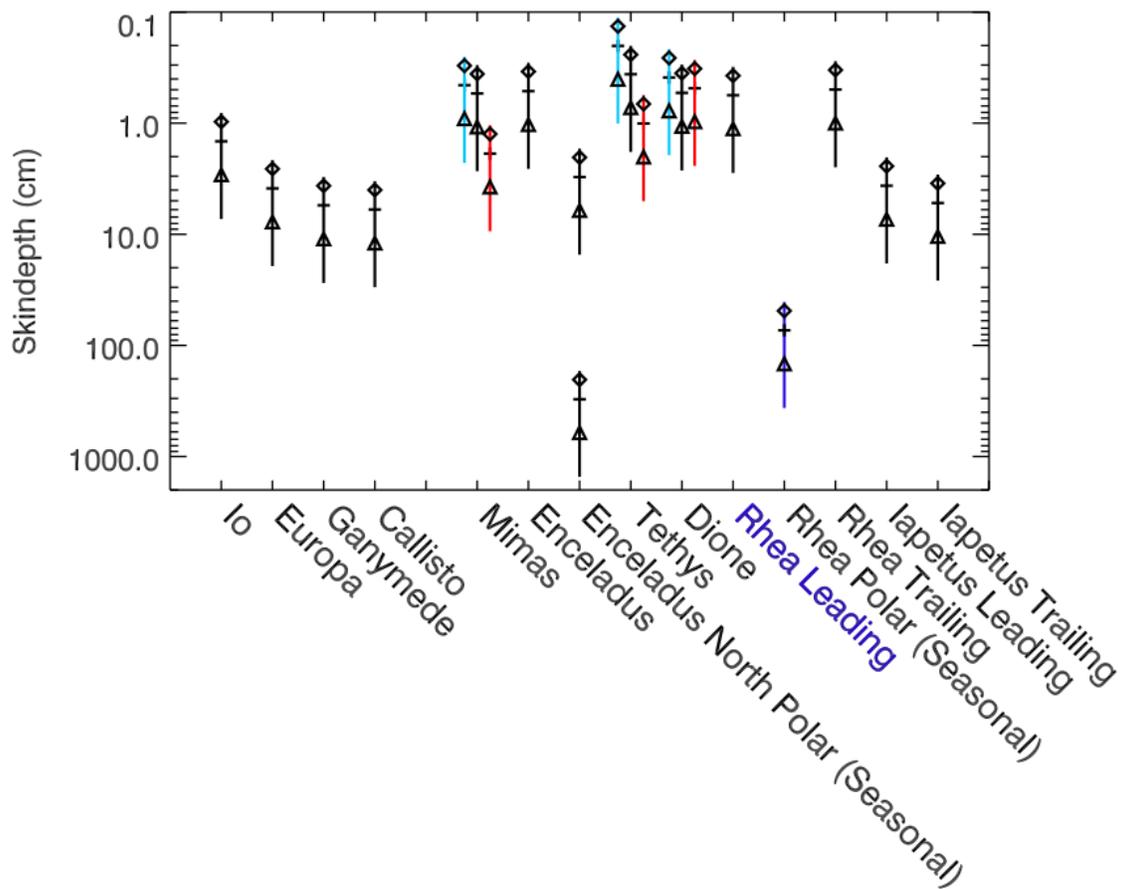

Figure 12 – The variation in skindepth across the major icy Saturnian satellites, determined using the thermal inertias provided in Figure 8. For Mimas, Tethys and Dione the blue and red lines indicate the thermal inertias outside and inside the thermally anomalous region respectively. The thermal inertia for Enceladus North Polar region as derived by Spencer *et al.* (2006) was an upper limit and therefore two bars are shown: the upper 100 MKS limit (deeper skindepth) and lower 1 MKS limit (shallower skindepth). For all targets the skindepths are given for a range of porosities (0.1 to 0.9), with diamonds, crosses and triangles symbols corresponding to a porosity of 0.25, 0.50 and 0.75 respectively. The same specific heat of water ice and zero porosity density as in Figure 9 is assumed, the specific heat of a basalt flow for Io's surface of 1.5 J K$^{-1}$ g$^{-1}$ (Davies,

1996). The skindepth of Rhea's polar region, as determined in this work, is shown in purple.

## 5 Acknowledgements

The authors would like to thank the Cassini Project and the Cassini Data Analysis Program (NNX12AC23G and NNX13AH84G) for funding this work.